# Higher-order Reconstruction Method of Differential Phase Shift

*Wenxiang Cong and Ge Wang*

*Biomedical Imaging Division, School of Biomedical Engineering and Sciences*

*Virginia Polytechnic Institute and State University, Blacksburg, VA 24061*

**Abstract**

In this paper, we develop a novel phase retrieval approach to reconstruct x-ray differential phase shift induced by an object. A primary advantage of our approach is a higher-order accuracy over that with the conventional linear approximation models, relaxing the current restriction of weak absorption and slow phase variation scenario. The optimal utilization of the diffraction images at different distance in Fresnel diffraction region eliminates the nonlinear terms in phase and attenuation, and simplifies the reconstruction to a linear inverse problem. Numerical studies are also described to demonstrate the accuracy and stability of our approach.

## 1. Introduction

Biological soft tissues encountered in clinical and pre-clinical imaging are mainly composed of atoms of light elements with low atomic numbers, and its elemental composition is nearly uniform with little density variations. The x-ray attenuation contrast is relatively poor, and cannot achieve satisfactory sensitivity and specificity [1, 2]. In contrast, x-ray phase-contrast provides a new mechanism for soft tissues imaging. The x-ray phase shift of soft tissues is about a thousand times greater than that of absorption within the diagnostic x-ray energy range, achieving a greater signal-to-noise ratio than attenuation contrast images. Thus, phase-contrast imaging can reveal detailed structural variation in soft tissues, offering a high contrast resolution between healthy and malignant tissues [3-5]. Moreover, the phase-contrast imaging does not intrinsically rely on x-ray absorption in tissues. With increasing x-ray energy, photoelectric absorption of tissues decreases as $1/E^3$, while tissue phase shift decreases much slower only as $1/E$. Phase imaging would significant reduce the deposited dose in tissues comparing to absorption contrast imaging within hard x-ray spectral range[6].

As it is well known, the x-ray intensity variation is acquired by a detector, while the phase shift of x-ray passing through an object cannot be measured directly. X-ray Talbot interferometry has recently been proposed as a novel x-ray phase imaging method to efficiently extract quantitative differential phase information from the Moiré pattern using a fringe scanning technique [4, 7]. However, the data acquisition procedure is quite time-consuming, resulting in an increasing radiation dose. The gratings with large sizes and high slit aspects are difficult to fabricate and model, especially, the analyzer absorption grating consists of Au pillars encased in epoxy and bounded using a frame. This fabrication process is prone to errors and hard to control in the case of gratings of large sizes and high aspects. X-ray phase contrast is formed from the propagation of wave field in free space after interaction with the object, creating a quantitative correspondence between the object and the recorded images, which can be used to retrieve the phase shift induced by the object. Several phase retrieval methods in the Fresnel diffraction regime are proposed, such as transport of intensity equation (TIE) method [8, 9], the contrast transfer function (CTF) [10], a mixed approach between the CTF and TIE [11], and a general theoretical formalism for the in-line phase-contrast imaging [12]. In this paper, we propose a novel method to reconstruct x-ray differential phase shift of an object. The reconstruction method keeps a higher-order accuracy comparing to the linearized approximation models, helping overcome the limitation in the weak absorption scenario and the restriction of slow phase variation assumed in the conventional phase reconstruction techniques. Using several intensity measurements in Fresnel diffraction region eliminate the nonlinear terms and simplify the phase retrieval to a linear model with respect to the differential phase shift.

**2. Phase retrieval algorithm**

The coherent x-ray-tissue interaction causes the x-ray wave phase change because of both x-ray diffraction and refraction effects. The amount of the phase change is determined by the dielectric susceptibility, or equivalently, by the refractive index of the tissue. The refractive index of x-ray is a complex form: $n = 1 - \delta + i\beta$, where the parameters $\delta$ and $\beta$ are the refractive index decrement and the absorption index, respectively. $\delta$ characterizes the x-ray phase shift, while $\beta$ is related to the attenuation properties. It has been shown that $\delta$ values ($10^{-6}$–$10^{-8}$) is about 1000 times greater than $\beta$ ($10^{-9}$– $10^{-11}$) in the biological soft tissue over the 10 keV-100 keV range [13]. This implies that tremendous improvement can be achieved in terms of the sensitivity of x-ray imaging to biological soft tissues if x-ray phase information is utilized. When an object is illuminated by a partially coherent x-ray beam, the wave-object interaction can be described as a transmittance function,

$$u_0(\mathbf{r}) = a(\mathbf{r})\exp[i\Phi(\mathbf{r})] = \exp[-B(\mathbf{r}) + i\Phi(\mathbf{r})] \tag{1}$$

where $\mathbf{r}$ denotes the transverse coordinates $(x, y)$ in the transverse plane relative to the propagation direction z. The absorption $B(\mathbf{r})$ and phase shift $\Phi(\mathbf{r})$ induced by the imaged object are the projections through the complex refractive index distribution. The phase change and attenuation of the x-ray beam after passing through an object are given by [14]

$$\Phi(\mathbf{r}) = \frac{2\pi}{\lambda}\int \delta(\mathbf{r}, z)dz, \quad B(\mathbf{r}) = \frac{4\pi}{\lambda}\int \beta(\mathbf{r}, z)dz \tag{2}$$

where $\lambda$ is the x-ray wavelength.

Due to phase shifts caused by an object, the wavefront of x-ray beam propagation is deformed. According to the Fresnel diffraction theory, the relationship between wave amplitudes in the transverse plane is described by the Fresnel transformation formula,

$$u_z(\mathbf{r}) = \frac{\exp(ikz)}{i\lambda z} u_0(\mathbf{r}) ** \exp(ik|\mathbf{r}|^2/2z) \tag{3}$$

where $k = 2\pi/\lambda$ and $**$ denotes convolution over the transverse coordinates $\mathbf{r}$. From Eq. (3), the Fourier transform of a Fresnel diffraction pattern can be written as [10],

$$\iint |u_z(\mathbf{r})|^2 \exp(i2\pi\mathbf{r}\cdot\mathbf{w})\mathbf{dr} = \iint A\left(\mathbf{r}-\frac{\lambda z}{2}\mathbf{w}\right) A\left(\mathbf{r}+\frac{\lambda z}{2}\mathbf{w}\right)\exp\left[i\Phi\left(\mathbf{r}+\frac{\lambda z}{2}\mathbf{w}\right)-i\Phi\left(\mathbf{r}-\frac{\lambda z}{2}\mathbf{w}\right)\right]\exp(i2\pi\mathbf{r}\cdot\mathbf{w})\mathbf{dr},$$
(4)

where $A(\mathbf{r})$ is the intensity distribution on the object plane. Because the hard x-ray wavelengths are very short in biomedical imaging and the near field setting, we have:

$$\Phi\left(\mathbf{r}+\frac{\lambda z}{2}\mathbf{w}\right)-\Phi\left(\mathbf{r}-\frac{\lambda z}{2}\mathbf{w}\right) \approx \lambda z \mathbf{w}\cdot\nabla\Phi(\mathbf{r})$$

$$A\left(\mathbf{r}-\frac{\lambda z}{2}\mathbf{w}\right)A\left(\mathbf{r}+\frac{\lambda z}{2}\mathbf{w}\right) \approx A^2(\mathbf{r})-\left(\frac{\lambda z}{2}\right)^2 (\mathbf{w}\nabla A(\mathbf{r}))^2$$
(5)

Taking the second-order approximation of the phase term, Eq. (4) at the image plane can be approximated by

$$\iint |u_z(\mathbf{r})|^2 \exp(i2\pi\mathbf{r}\cdot\mathbf{w})\mathbf{dr} = \iint\left[A^2(\mathbf{r})-\left(\frac{\lambda z}{2}\mathbf{w}\nabla A(\mathbf{r})\right)^2\right]\left[1+i\lambda z\mathbf{w}\cdot\nabla\Phi(\mathbf{r})-\frac{\lambda^2 z^2}{2}(\mathbf{w}\cdot\nabla\Phi)^2\right]\exp(i2\pi\mathbf{r}\cdot\mathbf{w})\mathbf{dr}$$
(6)

Using the measured intensity images at four different distances $z_i$ ($i=1,2,3,4$) from the object in Fresnel diffraction region, which corresponds to four equations based on Eq. (6). Furthermore, the high order term can be eliminated from the equation system, and the solution of phase retrieval is reduced to a linear integral equation with respect to the differential phase shift $\nabla\Phi$,

$$\iint_\Omega P(\mathbf{r})\exp(i2\pi\mathbf{r}\cdot\mathbf{w})dr = i\iint_\Omega A^2(\mathbf{r})\mathbf{w}\nabla\Phi\exp(i2\pi\mathbf{r}\cdot\mathbf{w})dr \tag{7}$$

where $P(\mathbf{r})=\sum_{i=1}^{4} t_i\left(|u_{z_i}|^2 - A^2\right)$ with weight coefficients $t_i = -2z_1 z_2 z_3 z_4 \Big/ \left[\lambda z_i^2 \prod_{i\neq j}(z_i - z_j)\right]$, $(i=1,2,3,4)$.

Eq. (7) is also equivalent to following differential equation,

$$\nabla(A^2(\mathbf{r})\nabla\Phi) = -2\pi\, P(\mathbf{r}) \tag{8}$$

which is a generalization of transport of intensity equation (TIE). TIE is valid in the limit of small object-to-detector distances. The differential phase shift $\nabla\Phi$ can be solved using the Fourier transform method based on Eq. (7) [15, 16]. Eq. (6) has been derived by taking advantage of a second-order approximation to the phase term, which helps to relax the limitation of weak absorption and slowly phase variation, making the differential phase reconstruction more accurate and stable. The phase shift of an x-ray beam passing through tissues is described as a projection, and can be used as input to a classic tomographic reconstruction algorithm to create a 3D reconstruction of the refractive index $\delta(\mathbf{r},z)$ in Eq. (2). Although the filtered backprojection (FBP) algorithm has been shown to be exact and fast implementation in the tomographic imaging [17], the iterative reconstruction approach also offers distinct advantages relative to analytic reconstruction in the cases of noisy and incomplete data. Especially, the recently emerging compressive sensing theory is powerful in solving under-determined inverse problems. Compressive sensing allows an accurate and stable image reconstruction using far fewer samples or measurement data than what are usually required, which can significantly reduce the number of projections, resulting in a significant reduction of radiation dose [18].

## 3. Numerical experiments

To demonstrate the feasibility of our proposed reconstruction approach, we carried out a series of numerical experiments using Beatle's phase and attenuation images, which come from the European Synchrotron Radiation Facility (ESRF, Grenoble). These images have a variety of features on different resolution scale, and consist of 812×744 pixels per image with a 5μm resolution. The radiation wavelength was set to λ=0.3Å which corresponds to hard x-rays. Using the Fresnel transform, we calculated the intensity distributions $|u_{z_i}|, (i=1, 2, 3, 4)$ on the image planes at propagation distances $z_1 = 200$, $z_2 = 350$, $z_3 = 500$, and $z_4 = 650$ mm, respectively, as shown in Fig. 1. Poisson noise was added to these intensity images to simulate the experimental condition. The proposed phase retrieval algorithm was applied to reconstruct the differential phase shift images. The reconstruction result is presented in Fig. 2 (b). We also implement a TIE-based reconstruction using the intensity image on object plane and an intensity image at a small propagation distance of 10m to the object. These results show that the proposed reconstruction method has a more high precision and stable than TIE method.

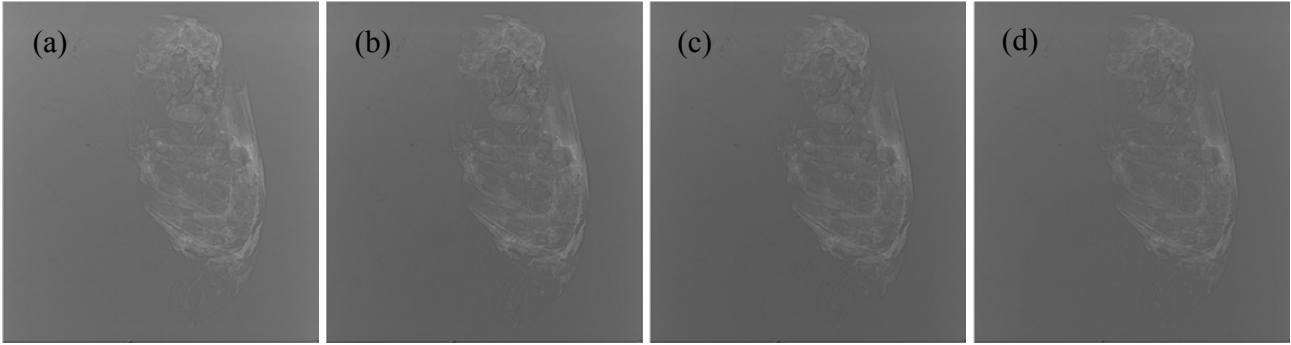

*Fig.1*. Intensity images at propagation distances 200 (a), 350 (b), 500 (c), and 650(d) mm, respectively.

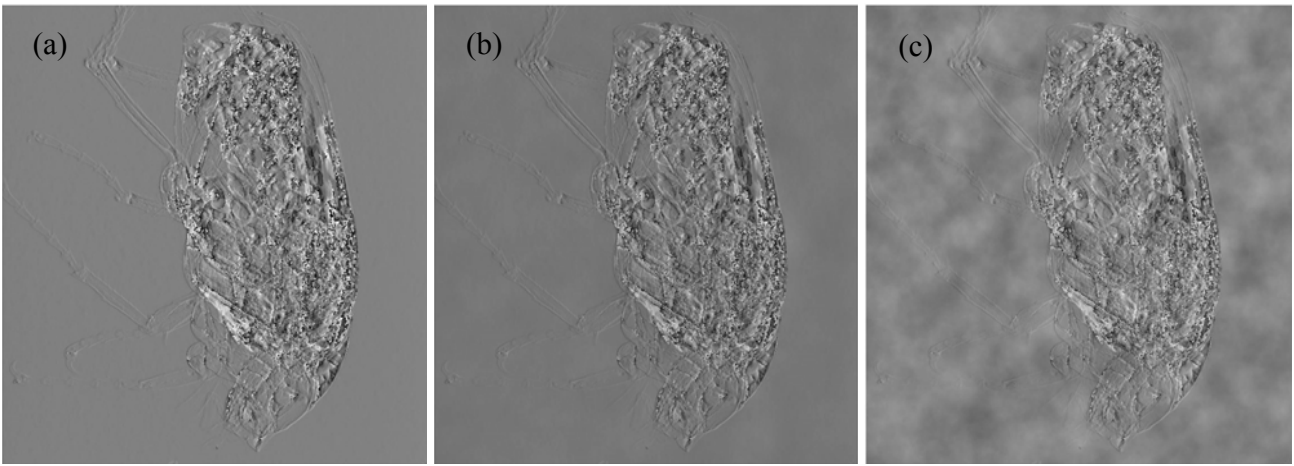

*Fig.2.* Comparison of reconstruction results. (a) Original phase image; (b) reconstructed phase image with the proposed method; (c) reconstructed phase image with the TIE method.

**4. Conclusion and discussions**

In summary, a novel reconstruction method is proposed for the retrieval of x-ray differential phase shift induced by an object. The numerical experiments demonstrated the accuracy and stableness of the reconstruction method for x-ray differential phase shift. The reconstruction method has a higher-order accuracy comparing to the linearized approximation models to overcome the limitation in the weak absorption scenario and the restriction of slow phase variation assumed in the conventional phase reconstruction techniques. The optimal utilization of the diffraction images at different distance eliminates the effect of nonlinear terms in phase and attenuation, and simplifies the reconstruction process to a linear inverse problem, outperforming the conventional iterative scheme of phase reconstruction [11] that the convergence remains an open problem. Theoretically, our method can be extended to a high-order approximation for the phase term at cost of intensity measurement at more distances. In the grating interferometer techniques, the complicated phase and analyzer absorption gratings is employed to extract the differential phase shift, being enslaved to the fabrication of grating with large sizes and high slit aspects. Our reconstruction approach retrieves the differential phase shift of an object only using the x-ray intensity measurement, significant simplifying the imaging system design and cost. It would replace the grating interferometer techniques to develop a new phase imaging system coupled with single source grating. In addition, the reconstruction formula (8) is also a generalization of transport of intensity equation (TIE). It has the second order computational accuracy, while TIE is only computed using a linear approximation, which is less robust to noise due to the lack of redundancy, related only two intensity images.